\documentclass[11pt]{article}

\usepackage{aas_macros,amsmath,amssymb,comment,cite,esint,graphicx,mathtools}
\usepackage[margin=.8in,letterpaper]{geometry}
\usepackage[colorlinks=true]{hyperref}
\usepackage[affil-it]{authblk}
\usepackage{subcaption}
\usepackage[utf8]{inputenc}
\usepackage{mathrsfs}
\usepackage{appendix}
\usepackage{amssymb}
\usepackage{float}
\usepackage{color}
\usepackage{cite}
\usepackage{hyperref}
\hypersetup{pageanchor=false}
\usepackage{indentfirst}
\usepackage{url}
\usepackage{float}
\usepackage{caption}
\usepackage[numbers,square,comma,sort&compress,merge]{natbib}
\usepackage{esint}
\usepackage{overpic}
\usepackage{graphicx}
\usepackage{epsf,amsmath,bbold,amsfonts,stmaryrd}
\usepackage{textcomp}
\usepackage{ulem}
\usepackage{tikz}

\numberwithin{equation}{section}
\let\originalleft\left
\let\originalright\right
\renewcommand{\left}{\mathopen{}\mathclose\bgroup\originalleft}
\renewcommand{\right}{\aftergroup\egroup\originalright}
\mathcode`\*="8000
{\catcode`\*=\active\gdef*{\mathclose{}\,\mathopen{}}}

\newcommand{\be}{\begin{equation}}
\newcommand{\ee}{\end{equation}}
\newcommand{\bea}{\setlength\arraycolsep{2pt} \begin{eqnarray}}
\newcommand{\eea}{\end{eqnarray}}

\begin{document}
\title{Energy extraction from a rotating black hole via magnetic reconnection: the plunging bulk plasma and orientation angle}

\author{
Ye Shen$^{1}$\thanks{E-mail: shenye199594@stu.pku.edu.cn},~
Ho-Yun YuChih$^{1}$\thanks{E-mail: hyyc@stu.pku.edu.cn}~,
Bin Chen$^{2,3,1,4}$\thanks{Corresponding author, E-mail: chenbin1@nbu.edu.cn} 
}
\date{}

\maketitle
\vspace{-10mm}

\begin{center}
	{\it	
      $^1$School of Physics, Peking University, No.5 Yiheyuan Rd, Beijing
		100871, P.R. China\\\vspace{4mm}

  $^2$Institute of Fundamental Physics and Quantum Technology, Ningbo University, Ningbo, Zhejiang 315211, China\\\vspace{4mm}

	$^3$School of Physical Science and Technology, Ningbo University, Ningbo, Zhejiang 315211, China\\\vspace{4mm}
		
		$^4$Center for High Energy Physics, Peking University,
		No.5 Yiheyuan Rd, Beijing 100871, P. R. China\\\vspace{4mm}
		
	}
\end{center}

\vspace{8mm}

\begin{abstract}
	\vspace{5mm}
Magnetic reconection provides a new avenue to extract energy from a rotating black hole in astrophysical scenarios. There have been many works studying the feasibility of extracting energy via magnetic reconnection in the scheme of magnetohydrodynamics. However,  most of them focused on circularly flowing bulk plasma only, and the influence of orientation angle, the angle between flowing direction of bulk plasma and ejection direction of plasmoids, was never carefully considered. In this work, we would like to study the energy extraction via magnetic reconnection, via the so-called Comisso-Asenjo process specifically, in the plunging plasma. More importantly, we analyze the influence of orientation angle in depth. Actually, we consider the magnetic reconnection occurring in bulk plasma with two kinds of streamlines, one being the circular flow and the other being the plunging flow from the ISCO. We find that the plunging plasma has higher  energy-extraction efficiency. Moreover, we notice that it is favorable for energy extraction when the orientation angle is suitably increased if the bulk plasma plunges. We further define the covering factor. Under the assumption that the magnetic reconnection occurs equally probable along any direction on any position within ergosphere, This value help us to primitively judge the probability of energy extraction via magnetic reconnection, through which we conclude that the plunging bulk plasma is much more capable in energy extraction than the circularly flowing one. 


\end{abstract}

\maketitle

\newpage
\baselineskip 18pt

\section{Introduction}
\label{sec:intro}

Magnetic reconnection, a widely known physical process to release magnetic energy efficiently, is believed to play a vital role in an amount of astrophysical phenomena. It was used to explain the solar and stellar flares \cite{substorm,2021SSRv..217...66Z}, the coronal mass ejection in the solar corona \cite{Lin-Forbes-2000} and the corona above the accretion disk of compact object \cite{Yuan2024-2}, the hot spots generated in the jetted flow \cite{Aimar2023,Ripperda2020} and the resultant episodic jets \cite{Yuan2009}, among others. It has been believed to occur in magnetized accretion flow near the massive black holes, especially on the equatorial plane of highly magnetized accretion system where the anti-parallel magnetic field lines form, move closer and get twisted under gravitational effect \cite{Yuan2024-1,Yuan2024-2,Ripperda:2021zpn}. As one of the most important astrophysical processes, the magnetic reconnection has been described by several analytical models  in the magnetohydrodynamic (MHD) scheme \cite{SP1,SP2,yamada2009,Petschek,Liu2017,Lyubarsky2006}, as well as by robust numerical simulations \cite{Spitkovsky:2005vsy,Sironi:2014jfa,Comisso:2023ygd} based on the particle-in-cell (PIC) method \cite{Birdsall:1981yj,Birdsall:1981yk}.

Since it is deemed to happen frequently in the magnetized accretion flows near the black holes, magnetic reconnection has been theoretically proposed as an mechanism of energy extraction from a rotating black hole, besides other mechanisms such as the superradiance of a massive bosonic field \cite{Brito:2015oca} and the Blandford-Znajek (BZ) mechanism \cite{BZ}. As we know, for a spinning black hole, a substantial fraction of its energy could be extracted such that the mass of black hole keeps being reduced until its irreducible value \cite{Christodoulou:1970wf}. Penrose envisioned a thought experiment that particle fission may occur within the ergo sphere of a rotating black hole, with one generated particle carrying negative energy as viewed from infinity while the other one carrying larger energy than that of the initial particle and escaping to infinity \cite{Penrose}. However, this so-called Penrose process suffers from the lack of initiation mechanism and has been believed to be hard to actualize in astrophysical scenario \cite{Wald:1974kya,Bardeen:1972fi}. Among many other mechanisms in extracting the energy from a spinning black hole, magnetic reconnection has drawn great attention  in recent years. It was Koide and Arai \cite{KA2008} who first explored the feasibility of energy extraction from a Kerr black hole. In their work, the ejected plasma was treated as relativistic adiabatic incompressible ball, the magnetic reconnection is treated as a slowly diffusive process and the magnetic field is set to be purely toroidally. Later on, Comisso and Asenjo creatively investigated the dynamics of expelled plasmoid \cite{CA2021}, produced by fast magnetic reconnection within magnetized bulk plasma \cite{Sironi:2022hnw,French:2022zfv}, and found that the energy extraction via magnetic reconnection could be more efficient than the one via the BZ mechanism. The energy extraction from a rotating black hole via magnetic reconnection, via the so-called Comisso-Asenjo process specifically, has been further explored in various spacetimes \cite{Carleo:2022qlv,Wei:2022jbi,Liu:2022qnr,Wang:2022qmg,Zhang:2024ptp,Rodriguez:2024jzw}.

Most of previous works about the Comisso-Asenjo process focused on the circularly flowing bulk plasma on the equatorial plane around the central black hole. However, it was proposed that plasma within the innermost stable circular orbit (ISCO) flows along the plunging geodesic (called intra-ISCO inspiral initially) down to the central black hole \cite{Mummery2022,Mummery2023,Mummery2024}, since circular orbits within the ISCO are unstable. As the plunging plasma may have significantly different dynamics from those in circular accretion flow, it is essential to study the magnetic reconnection within the ISCO, the so-called plunging region \cite{Wilkins:2020pgu,Dong:2023bbd}. In Ref.~\cite{Work0}, the authors primitively analyzed the probability of energy extraction via the Comisso-Asenjo process from a Kerr black hole in the plunging region, by assuming the plunging bulk plasma move on the equatorial plane. In the study of Ref.~\cite{Work0}, the direction of plasmoid ejection was fixed in order for the large scale structure of magnetic field strictly obeying the ideal MHD condition. However,  magnetic field configurations are  complicated and protean in real astrophysical systems, which ultimately results in an arbitrary direction of plasmoid ejection \cite{CA2021}. In this study, we promote the analyses in Ref.~\cite{Work0} by relaxing the requirement on the orientation angle, the angle between the direction of accretion flow and the ejection direction of plasmoids. More importantly, we compare the efficiencies of energy extraction via the Comisso-Asenjo process occurring in the plunging and the circularly flowing bulk plasma, under the differently chosen orientation angles. Furthermore, we compare the allowed regions for energy extraction in two kinds of bulk plasma in several parameter planes. We will see from the figures and basic analyses that the energy extraction in the plunging bulk plasma is more efficient than in the circularly flowing bulk plasma. Besides, the influence of orientation is no longer trivial if the bulk plasma plunges, from the perspective of which we figure out the best orientation angle in the high local-magnetization limit. In order to primitively study how probable the energy extraction would succeed in an accretion system by assuming that the magnetic reconnection occurs equally probable in any direction on any position of the ergo region, we define a quantity $\chi$, named the covering factor, to be the ratio of area of the allowed region for energy extraction to the total area in the parameter plane of reconnection point and orientation angle. By solving the equation $\chi=0$, we are allowed to determine the minimal spin of the black hole to allow energy extraction for a fixed magnetization, and to determine  the minimal value of the local magnetization for a fixed spin. We show that contrary to the circularly flowing bulk plasma, it is still possible to extract the energy from a slowly spinning black hole if the bulk plasma plunges within the  ISCO.

The remain parts of the paper are organized as follows. We will briefly discuss basic concepts in Sect.~\ref{sec:cal}, including the metric and normal tetrads, the description and trajectories of bulk plasma, and the information of the Comisso-Asenjo process. In Sect.~\ref{sec:efficiency}, we will compare the efficiencies of and allowed regions for the energy extraction via magnetic reconnection occurring in the plunging and circularly flowing bulk plasma with zero orientation angle. In Sect.~\ref{sec:angle}, we will further consider the dependence of energy extraction on the orientation angle and plot the allowed regions in the parameter plane consisting of reconnection point and the orientation angle. In Sect.~\ref{sec:chi}, we will define the covering factor for energy extraction, through which we will find out the minimal allowed local magnetization for energy extraction with a fixed black spin or on the contrary the minimal black hole spin with local magnetization being fixed. We summarize our work in Sect.~\ref{sec:sum}. We adopt the units $G=M=c=1$ throughout the paper.

\section{Bulk plasma and ejection of plasmoids}
\label{sec:cal}


Let us begin with revisiting some basic concepts of bulk plasma and Comisso-Asenjo process in a stationary, axisymmetric spacetime, which was initially introduced in Ref.~\cite{CA2021}. It is effective to represent the equations in 3+1 formalism \cite{MacDonald:1982zz}, in which the line element can be written as
\begin{equation}
    ds^2=g_{\mu\nu}dx^{\mu}dx^{\nu}=-\alpha^2dt^2+\sum_{i=1}^3\left(h_idx^i-\alpha\beta^idt\right)^2
    \label{eq:line}
\end{equation}
where $\alpha$ is the lapse function,  $\beta^i=h_i\omega^i/\alpha$ is the shift vector with $\omega^i$ being the velocity of frame dragging, and $h_i$ being the scale factor. In this work, we focus on the spacetime given by the Kerr metric in the BL coordinates $\left(t,r,\theta,\phi\right)$, for which
\begin{equation}
    \alpha=\sqrt{\frac{\Delta\Sigma}{A}},~~h_r=\sqrt{\frac{\Sigma}{\Delta}},~~h_{\theta}=\sqrt{\Sigma},~~h_{\phi}=\sqrt{\frac{A}{\Sigma}}\sin\theta,~~
    \omega^r=\omega^{\theta}=0,~~\omega^{\phi}=\frac{2ar}{A}
    \label{eq:alpha_h_omega}.
\end{equation}
Here $\Sigma=r^2+a^2\cos^2\theta$, $\Delta=r^2-2r+a^2$ and $A=\left(r^2+a^2\right)^2-a^2\Delta\sin^2\theta$. 

The zero-angular-momentum-observers (ZAMOs) in the spacetime could be defined via the normal tetrad, which is of the form
\begin{equation}
    \hat{e}_{(t)}^{\mu}=\frac{1}{\alpha}\left(\partial_t^{\mu}+\omega^{\phi}\partial_{\phi}^{\mu}\right),~~
    \hat{e}_{(r)}^{\mu}=\frac{1}{h_r}\partial_r^{\mu},~~~~
    \hat{e}_{(\theta)}^{\mu}=\frac{1}{h_{\theta}}\partial_{\theta}^{\mu},~~
    \hat{e}_{(\phi)}^{\mu}=\frac{1}{h_{\phi}}\partial_{\phi}^{\mu}.
    \label{eq:ZAMOs}
\end{equation}
The magnetic reconnection is assumed to happen in the bulk plasma (described by perfect fluid approximately) moving on the equatorial plane. It is necessary to quantify the magnetic reconnection in the fluid's rest frame which can be defined via the normal tetrad as
\begin{equation}
    \begin{split}
        e_{[0]}^{\mu}&=\hat{\gamma}_s \left[\hat{e}_{(t)}^{\mu}+\hat{v}_s^{(r)}\hat{e}_{(r)}^{\mu}+\hat{v}_s^{(\phi)}\hat{e}_{(\phi)}^{\mu}\right], \\
        e_{[1]}^{\mu}&=\frac{1}{\hat{v}_s}\left[\hat{v}_s^{(\phi)}\hat{e}_{(r)}^{\mu}-\hat{v}_s^{(r)}\hat{e}_{(\phi)}^{\mu}\right],~~
        e_{[2]}^{\mu}=\hat{e}_{(\theta)}^{\mu}, \\
        e_{[3]}^{\mu}&=\hat{\gamma}_s\left[\hat{v}_s\hat{e}_{(t)}^{\mu}+\frac{\hat{v}_s^{(r)}}{\hat{v}_s}\hat{e}_{(r)}^{\mu}+\frac{\hat{v}_s^{(\phi)}}{\hat{v}_s}\hat{e}_{(\phi)}^{\mu}\right]
    \end{split}
    \label{eq:fluid}
\end{equation}
where $\hat{v}_s=\sqrt{\left(\hat{v}_s^{(r)}\right)^2+\left(\hat{v}_s^{(\phi)}\right)^2}$ is the speed of fluid with $\hat{v}_s^{(r)}$ and $\hat{v}_s^{(\phi)}$ the components of 3-velocity, and $\hat{\gamma}_s$ is the Lorentz factor, observed by the ZAMOs. It is clear that, viewed by the ZAMOs, $e_{[1]}$ and $e_{[3]}$ are orthogonal and parallel to the fluid's moving direction, respectively.


It seems that at the horizon scale ($r\simeq 1$), the plasma would be nearly collisionless and should be treated in the framework of Vlasov gas dynamics \cite{Vlasov1968}. However, for simplicity, we still employ the perfect fluid approximation to describe the plasma near the event horizon. The energy conservation, $\nabla_{\mu}T^{\mu}_{~t}=0$, leads to
\begin{equation}
    \partial_te+\frac{1}{h_rh_{\theta}h_{\phi}}\partial_{i}\left(h_rh_{\theta}h_{\phi}S^i\right)=0,
    \label{eq:energy_conserve}
\end{equation}
where $e=-\alpha T^t_{~t}$ is the energy-at-infinity density and $S^i=\alpha T^i_{~t}$ is the flux density. The stress-energy tensor of plasma could be separated into fluid part and electromagnetic part as $T^{\mu\nu}=T_{\rm F}^{\mu\nu}+T_{\rm EM}^{\mu\nu}$, where:
\begin{equation}
    T_{\rm F}^{\mu\nu}=\omega u^{\mu}u^{\nu}+pg^{\mu\nu}~~~~,~~~~
    T_{\rm EM}^{\mu\nu}=F^{\mu}_{~\kappa}F^{\kappa\nu}-\frac{1}{4}g^{\mu\nu}F_{\alpha\beta}F^{\alpha\beta}
    \label{eq:Tmn}
\end{equation}
with $p$, $\omega$ being the proper pressure and enthalty respectively, $u^{\mu}$ being the 4-velocity and $F^{\mu\nu}$ being the Maxwell's tensor of plasma.

The bulk plasma, in which the magnetic reconnection occurs, can be regarded to be perfect fluid flowing on the equatorial plane around the central black hole. Since the gravitational effect is dominant near the black hole, the fluid element could be regarded to move along timelike geodesics \cite{Hou:2023bep}, such that $U_{\mu}dx^{\mu}=-Edt+U_r(r)dr+Ld\phi$, where $U^r$ represents the 4-velocity of bulk plasma, $E$ and $L$ represent the conserved energy and angular momentum (per unit mass) of timelike geodesic in a Kerr spacetime. We can project $U^{\mu}$ onto the normal tetrad of ZAMOs as
\begin{equation}
    U^{\mu}\hat{e}^{(a)}_{\mu}=\hat{\gamma}_s\left\{1,\hat{v}_s^{(r)},0,\hat{v}_s^{(\phi)}\right\}=
    \left\{\frac{E-\omega^{\phi}L}{\alpha},h_rU^r,0,\frac{L}{h_{\phi}}\right\}.
    \label{eq:U}
\end{equation}
Two kinds of geodesics are under consideration. The first one is circular orbits (prograde only) on which $U^r=0$ and
\begin{equation}
    E=E_{\rm K}(r)=\frac{r^{3/2}-2r^{1/2}+a}{\sqrt{r^3-3r^2+2ar^{3/2}}},~~
    L=L_{\rm K}(r)=\frac{r^2-2ar^{1/2}+a^2}{\sqrt{r^3-3r^2+2ar^{3/2}}}.
    \label{eq:E_L}
\end{equation}
These circular orbits may exist from infinity down to the photon sphere whose radius is
\begin{equation}
    r_{\rm ph}=2\left[1+\cos\left(\frac{2}{3}\arccos(-a)\right)\right].
\end{equation}
The photon sphere is outside the ergo-sphere provided that $a>1/\sqrt{2}\simeq 0.7$. The other one is the plunging geodesic (prograde only as well) \cite{Mummery2022}, since the circular bulk plasma within the ISCO is unstable under a tiny perturbation and plunges from the ISCO. The radius of the ISCO is
\begin{equation}
    r_{\rm ms}=3+Z_2-\sqrt{\left(3-Z_1\right)\left(3+Z_1+2Z_2\right)}
    \label{eq:ISCO}
\end{equation}
where
\begin{equation}
    Z_1=1+\left(1-a^2\right)^{1/3}\left[\left(1+a\right)^{1/3}+\left(1-a\right)^{1/3}\right],~~
    Z_2=\sqrt{3a^2+Z_1^2}.
\end{equation}
The region for $r<r_{\rm ms}$ is called the plunging region \cite{Wilkins:2020pgu,Dong:2023bbd}, the circular orbits in which are unstable. In general, we have $r_{\rm EH}\leq r_{\rm ph} \leq r_{\rm ms}$, where $r_{\rm EH}$ is the radius of the event horizon. For a extremely spinning black hole ($a\simeq 1$) we have $r_{\rm ms}\simeq r_{\rm ph}\simeq r_{\rm EH}\simeq 1$. The conserved quantities of plunging geodesic are just set to be those of circular orbit on the ISCO:
\begin{equation}
    E_{\rm ms}=E_{\rm K}\left(r_{\rm ms}\right),~~L_{\rm ms}=L_{\rm K}\left(r_{\rm ms}\right),
    \label{eq:E_L_plunge}
\end{equation}
and in this case the radial component of 4-velocity takes a simple form of \cite{Mummery2022}
\begin{equation}
    U^r=-\sqrt{\frac{2}{3r_{\rm ms}}}\left(\frac{r_{\rm ms}}{r}-1\right)^{3/2}.
    \label{eq:Ur_plunge}
\end{equation} 

In this work, we consider two kinds of streamlines of bulk plasma. In the first case, the bulk plasma rotates circularly around the central black hole outside the ISCO before it plunges into the central black hole in the plunging region. This kind of streamline will be referred to as the combined streamline hereafter. In the second case, the bulk plasma always rotates in circular orbits with radii larger than the one of the photon sphere. Such kind of streamline will be called the circular streamline hereafter.  By applying Eq.~\eqref{eq:E_L}, \eqref{eq:E_L_plunge} and \eqref{eq:Ur_plunge} to Eq.~\eqref{eq:U}, we get the velocities of bulk plasma observed by the ZAMOs for both kinds of the streamlines. We will see the great differences in energy extraction via magnetic reconnection between the two cases in Sect.~\ref{sec:result}.


\begin{figure}
    \centering
    \includegraphics[width=0.9\textwidth]{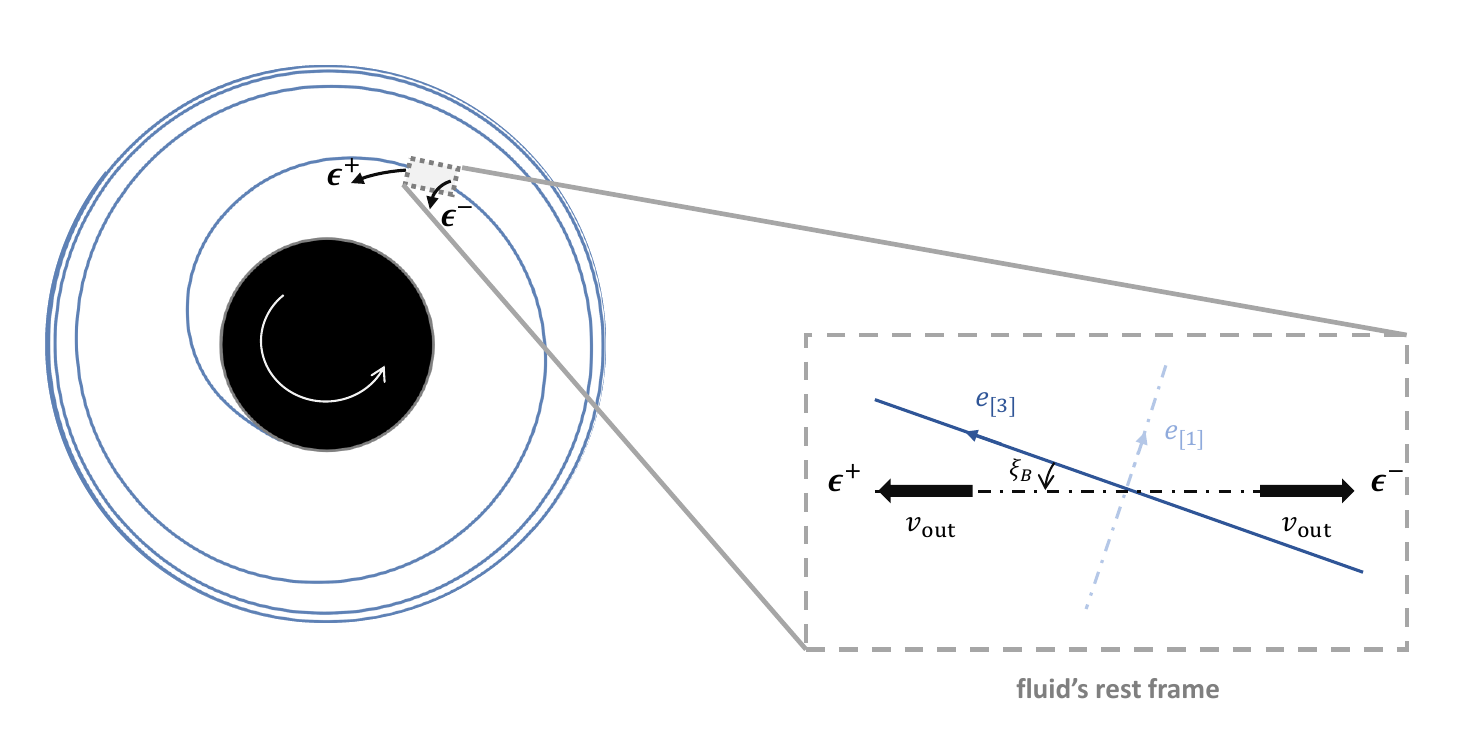}
    \caption{Left: Schematic diagram of the energy extraction via magnetic reconnection process, during which the plasmoids are ejected oppositely (black arrows), viewed in a large scale; The blue line represent the streamline of the bulk plasma. Right (grey dashed square): Details of the ejection of plasmoids at the reconnection point, view in the fluid's rest frame.}
    \label{fig:illu}
\end{figure}

Numerical simulations of MHD indicated that the rotation of black hole may produce antiparallel magnetic field lines adjacent to the equatorial plane \cite{Ripperda2020,Yuan2024-2}. The flip of magnetic field direction makes the existence of an equatorial current sheet possible. The equilibirium of anti-parallel magnetic field would be broken by plasmoid instability which may be induced by thermal-inertial effects, electric resistivity and so on \cite{Comisso:2016pyg,Comisso:2017arh,PhysRevLett.121.165101}. The plasmoid instability actuates fast magnetic reconnection \cite{Daughton2009}, which eventually converts magnetic energy into kinetic energy of plasma and then expels created plasmoids away. Upon the plasmoid instability occurring near the black hole, as introduced in Ref.~\cite{CA2021}, the Comisso-Asenjo process assumed two plasmoids ejected oppositely in the local rest frame of bulk plasma. The 4-velocites of the ejected plasmoids could be written as \cite{Work0}:
\begin{equation}
    u^{\mu}_{\pm}=\gamma_{\rm out}\left[e_{[0],0}^{\mu}\pm 
    v_{\rm out}\left(\cos\xi_{B}e_{[3],0}^{\mu}+\sin\xi_{B}e_{[1],0}^{\mu}\right)\right]
    \label{eq:u_out}
\end{equation}
with $v_{\rm out}$ being the outflow speed of plasmoids measured in the fluid's rest frame and $\gamma_{\rm out}$ being the Lorentz factor. The subscript "0" represents the reconnection point, the location where the magnetic reconnection happens. The "$\pm$" denotes two plasmoids ejected towards opposite directions. We adopt the theoretical model of fast magnetic reconnection described in Ref.~\cite{Liu2017}, which provides \cite{Work0}
\begin{equation}
    v_{\rm out}\simeq \sqrt{\frac{\left(1-{\sf g}^2\right)^3\sigma_0}{\left(1+{\sf g}^2\right)^2+\left(1-{\sf g}^2\right)^3\sigma_0}}
    \label{eq:v_out}
\end{equation}
where $\sigma_0$ represents the local magnetization of bulk plasma on the reconnection point and ${\sf g}$ is called the geometric index defined to be the ratio of the width to the length of current sheet. In this work, we choose ${\sf g}\simeq 0.49$ with which the local reconnection rate peaks in the case of high magnetization limit ($\sigma_0\gg 1$) \cite{Work0}.

The $\xi_{B}$ in Eq.~\eqref{eq:u_out} represents the orientation angle of ejected plasmoids, which is shown in the dashed square of Fig.~\ref{fig:illu}. Since the magnetic reconnection are generally described by 2D models \cite{yamada2009,Liu2017}, the orientation angle should just be the angle between $e_{[3]}^{\mu}$ and the direction of magnetic field lines at the reconnection point. In most previous works, only some values of $\xi_B$ were chosen in the analysis \cite{CA2021,Carleo:2022qlv}, mainly because only the circularly flowing bulk plasma was considered,  in which case it is easy to conclude that a non-zero orientation angle makes the energy extraction harder.  In Ref.~\cite{Work0},  the magnetic reconnection in a plunging plasma was studied, in which case $\xi_B$ was fixed based on ideal MHD condition $F_{\mu\nu}U^{\nu}=0$ in bulk plasma. In reality, however, magnetic field lines cannot be determined easily. The turbulence, induced by various complicated processes such as magnetic rotational instability \cite{MRI1,MRI2} in the accretion flow would make the magnetic field configuration twisted, unpredictable and drastically time-dependent. In this work, we regard $\xi_B$ to be a free parameter ranging from  $-\pi/2$ to $\pi/2$\footnote{The case with, for example, $\pi/2<\xi_B<\pi$ can be reached by making $\xi_B\rightarrow\xi_B-\pi$ and $\epsilon_+\leftrightarrow\epsilon_-$.} and consider its influence on the energy extraction. We will see in Sect.~\ref{sec:angle} that $\xi_B$ has great impact on the energy extraction, especially in the bulk plasma with the combined streamline.

Aligned with the convention in Ref.~\cite{CA2021}, the  energy-extraction efficiency is defined to be
\begin{equation}
    \eta=\frac{\epsilon_+}{\epsilon_++\epsilon_-}
    \label{eq:eta}
\end{equation}
where $\epsilon_{\pm}$ are the energy-at-infinity per enthalpy of the two ejected plasmoids. In the Comisso-Asenjo process, magnetic reconnection was assumed to be highly efficient such that most of the magnetic energy is converted into kenetic energy. In this sense, by applying the expressions of $T_{\rm F}^{\mu\nu}$ in Eq.~\eqref{eq:Tmn} and $u_{\pm}$ in Eq.~\eqref{eq:u_out}, we get \cite{CA2021,Work0}
\begin{equation}
    \begin{split}
        \epsilon_{\pm}=&\frac{e_{\pm}}{\alpha u^t_{\pm}\omega_{\pm}}=-u_t^{\pm}-\frac{\Tilde{p}_{\pm}}{u^t_{\pm}} \\
        =&\alpha\hat{\gamma}_s\gamma_{\rm out}\left[\left(1+\beta^{\phi}\hat{v}_s^{(\phi)}\right)\pm 
        v_{\rm out}\left(\hat{v}_s+\beta^{\phi}\frac{\hat{v}_s^{(\phi)}}{\hat{v}_s}\right)\cos\xi_B\mp
        v_{\rm out}\beta^{\phi}\frac{\hat{v}_s^{(r)}}{\hat{\gamma}_s\hat{v}_s}\sin\xi_B\right] \\
        &-\frac{\alpha\Tilde{p}}{\hat{\gamma}_s\gamma_{\rm out}\left(1\pm \hat{v}_s v_{\rm out}\cos\xi_B\right)}
    \end{split}
    \label{eq:epsilon}
\end{equation}
where $\Tilde{p}_{\pm}$ is the proper pressure per enthalpy of the plasmoids. We adopt $\Tilde{p}_{\pm}\simeq 1/4$ in this work by regarding the plasmoids to be relativistically hot \cite{Lyubarsky2006}. More details about $\epsilon_{\pm}$ could be found in Sect.~2.4 in Ref.~\cite{Work0}. A successful energy extraction via magnetic reconnection should generate a plasmoid with negative $\epsilon_-$, namely the efficiency $\eta$ should be greater than 1.

\section{Results}
\label{sec:result}

In this section, we present the results of our study. We compare the efficiencies of and the allowed regions for energy extraction via magnetic reconnection occurring in the plunging and the circularly flowing bulk plasma with zero orientation angle. Moreover we discuss  the dependence of energy extraction on the orientation angle and plot the allowed regions in the parameter plane consisting of reconnection point and the orientation angle. Finally, we define the covering factor $\chi$ to describe the probability of energy extraction under the assumption that the occurrences of magnetic reconnection in any direction on any position in ergo region are equally probable. By solving $\chi=0$ we determine the requirement on the spin and local magnetization for energy extraction via magnetic reconnections. 

\subsection{Efficiencies of energy extractions in two different streamlines}
\label{sec:efficiency}

\begin{figure}
    \centering
    \hspace{-2.5mm}
    \includegraphics[width=0.48\textwidth]{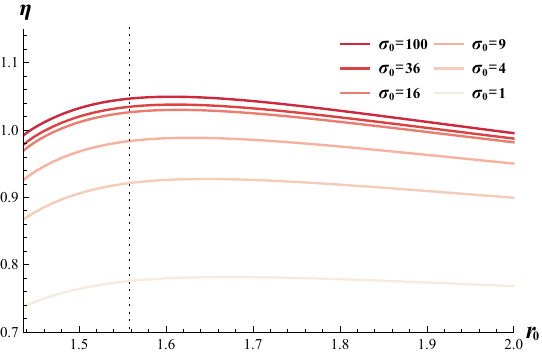}
    \hspace{2.5mm}
    \includegraphics[width=0.48\textwidth]{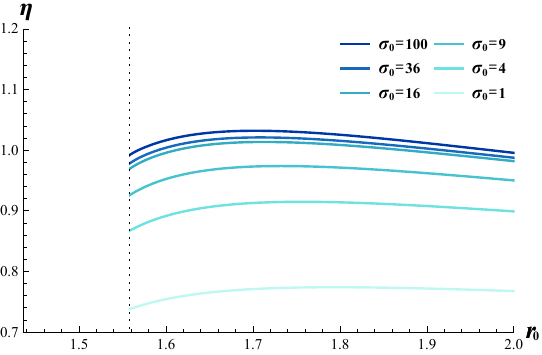}
    \vspace{5mm}
    \hspace{-2.5mm}
    \includegraphics[width=0.48\textwidth]{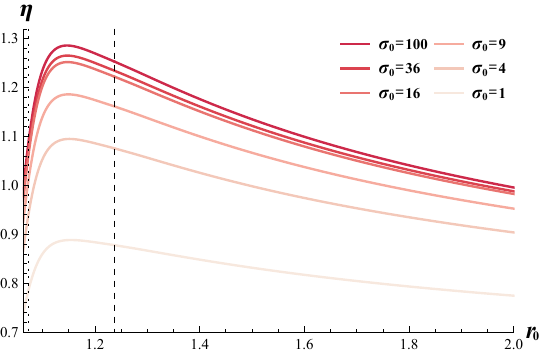}
    \hspace{2.5mm}
    \includegraphics[width=0.48\textwidth]{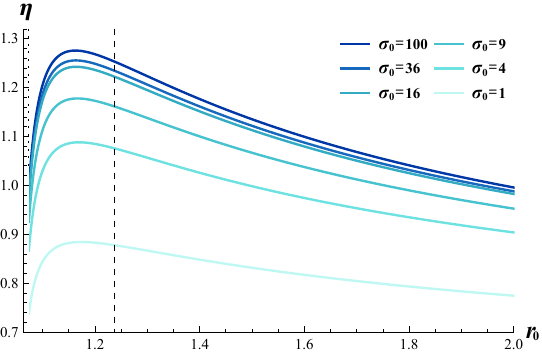}
    \caption{ energy-extraction efficiency $\eta$ as a function of the radius of the reconnection point $r_0$ for multiple values of local magnetization $\sigma_0$ with $\xi_B=0$, $a=0.9$ (top panels) and $a=0.998$ (bottom panels) in the cases of the combined streamlines (left panels) and the circular streamlines (right panels). The dotted lines represent the radii of photon spheres while the dashed lines represent the ISCO ($r_{\rm ms}>r_{\rm ergo}$ for $a=0.9$).}
    \label{fig:eta-sig}
\end{figure}

\begin{figure}
    \centering
    \hspace{-2.5mm}
    \includegraphics[width=0.48\textwidth]{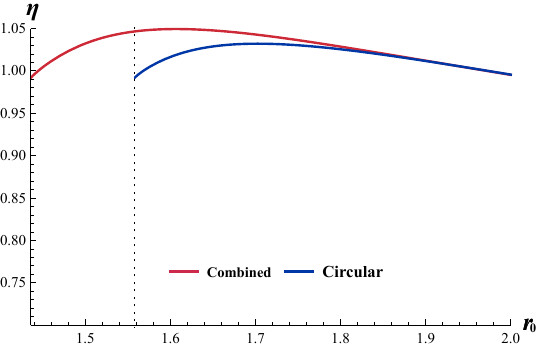}
    \hspace{2.5mm}
    \includegraphics[width=0.48\textwidth]{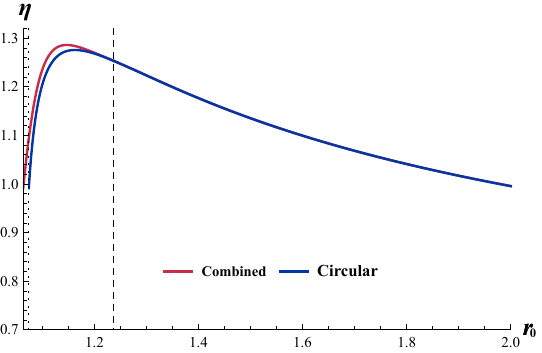}
    \caption{ Energy-extraction efficiency $\eta$ as a function of the radius of the reconnection point $r_0$ with $\xi_B=0$, $\sigma_0=100$, $a=0.9$ (left panel) and $a=0.998$ (right panel) in the cases of the combined streamlines (red lines) and the circular streamlines (blue lines). The dotted lines represent the radii of the photon sphere while the dashed lines represent the ISCO ($r_{\rm ms}>r_{\rm ergo}$ for $a=0.9$).}
    \label{fig:eta-ck}
\end{figure}

The energy-extraction efficiency, defined in Eq.~\eqref{eq:eta}, as a function of the radius of reconnection point $r_0$ is plotted in Fig.~\ref{fig:eta-sig} for multiple values of local magnetization in the cases of combined streamline (left panels) and circular streamline (right panels) with $\xi_B=0$, $a=0.9$ (top panels) and $a=0.998$ (bottom panels), where $r_0$ ranges from the radius of the event horizon $r_{\rm EH}$ to the radius of the ergo-sphere $r_{\rm ergo}$ on the equatorial plane. The radii of the photon sphere, where the circular orbits disappear, and the radii of  the ISCO are (henceforth) represented by the dotted and dashed lines respectively (no dashed line in the top panals because $r_{\rm ms}>2=r_{\rm ergo}$ when $a=0.9$). We can see from Fig.~\ref{fig:eta-sig} that the  energy-extraction efficiency increases slower and slower with the increase of local magnetization. It is consistent with the result in Ref~\cite{CA2021} and was also mentioned in Ref~\cite{Work0} that the maximal value of $\eta$ remains nearly constant when the local magnetization is high enough ($\sigma_0\gtrsim 50$). Actually, Eq.~\eqref{eq:v_out} indicates that $v_{\rm out}$ approaches 1 while $\gamma_{\rm out}$ is proportional to $\sqrt{\sigma_0}$, as the local magnetization approaches infinity. Applying those into Eq.~\eqref{eq:epsilon}, we have $\epsilon_{\pm}\propto\sqrt{\sigma_0}$ such that $\eta$ approximately keeps constant with the increase of $\sigma_0$ when $\sigma_0\gg 1$. It is consistent with the result shown in Eq.~(38) and Eq.~(39) of Ref.~\cite{CA2021}.

The efficiency, when $\xi_B=0$, in the case of combined streamline is generally higher than the one in the case of circular streamline in the plunging region, which could  be more easily  concluded from Fig.~\ref{fig:eta-ck}. It consequently suggests that the energy extraction could be more probable to happen if the bulk plasma plunges to the central black hole than still flows circularly in the plunging region, when the plasmoids are ejected along the moving direction of bulk plasma.

\begin{figure}
    \centering
    \hspace{-3mm}
    \includegraphics[width=0.48\textwidth]{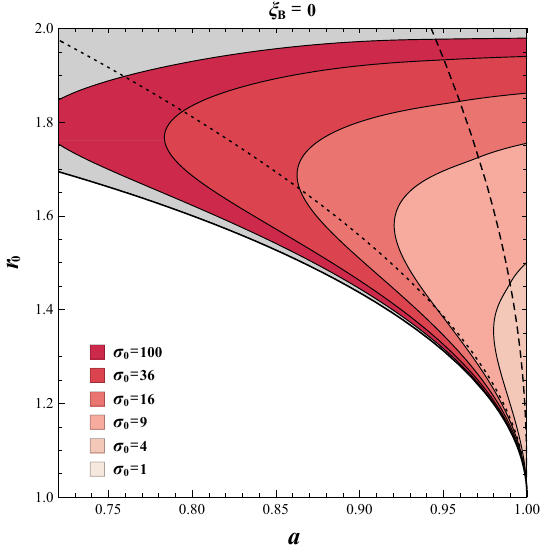}
    \hspace{3mm}
    \includegraphics[width=0.48\textwidth]{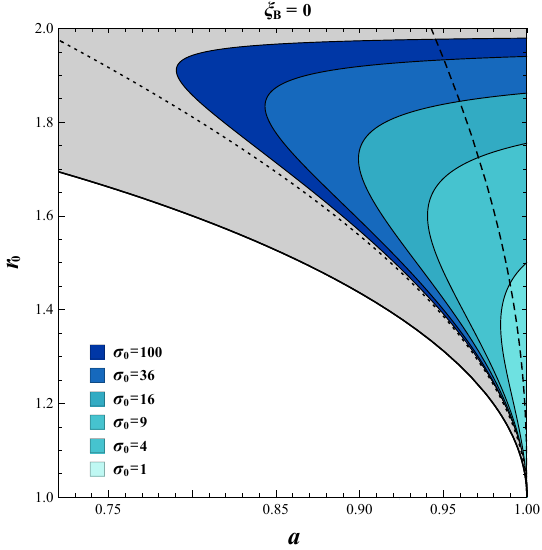}
    \caption{The $a-r_0$ planes in the cases of the combined streamline (left panels) and the circular streamline (right panels).The allowed regions for energy extraction ($\eta>0$), with $\xi_{\rm B}=0$, are colored in red (combined streamline) or blue (circular streamline) for multiple values of local magnetization $\sigma_0$. The solid, dotted and dashed black lines represent the radii of the event horizon, the photon sphere and the ISCO, respectively.}
    \label{fig:r-a-0}
\end{figure}

We plot the allowed regions for energy extraction (the region for $\eta>1$) on $a-r_0$ planes with $\xi_B=0$ for multiple values of local magnetization in Fig.~\ref{fig:r-a-0}. The allowed regions for energy extraction in the case of the combined streamline are (henceforth) colored in red while the one in the case of the circular streamline are (henceforth) colored in blue. The event horizons are (henceforth) represented by solid lines. Enhancing the local magnetization would obviously expand the allowed region for energy extraction. However, when the local magnetization is high enough ($\sigma_0\gtrsim 50$), the expansion is rather limited, which was concluded in Ref.~\cite{CA2021} foremost in the case of circular streamline only and is consistent with the conclusion found from Fig.~\ref{fig:eta-sig} as well.

We could also conclude from Fig.~\ref{fig:r-a-0} that the energy extraction from the bulk plasma with Keplerian streamline is more restrictive. It is partly because the circular orbits disappear within the photon sphere. In contrast, we can see from the left panel in Fig.~\ref{fig:r-a-0} that, within the photon sphere, there are still a wide region which is allowed for the energy extraction in the case of combined streamline. 

\subsection{The implications of non-vanishing orientation angle}
\label{sec:angle}

\begin{figure}
    \centering
    \hspace{-2.5mm}
    \includegraphics[width=0.48\textwidth]{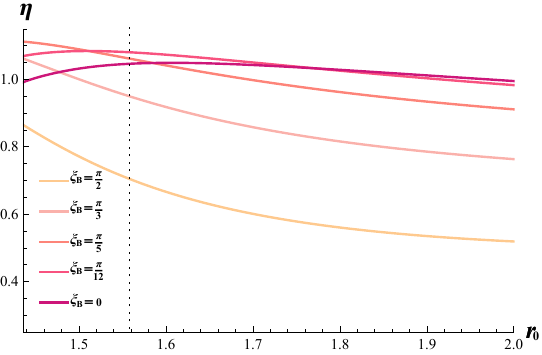}
    \hspace{2.5mm}
    \includegraphics[width=0.48\textwidth]{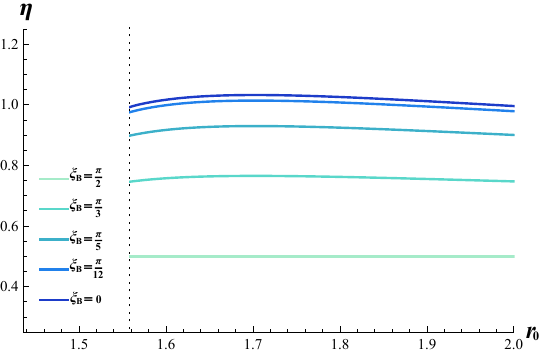}
    \hspace{-2.5mm}
    \includegraphics[width=0.48\textwidth]{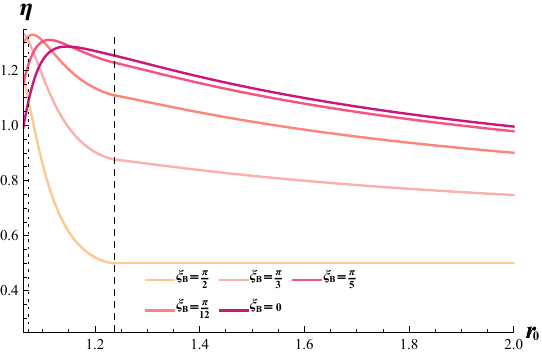}
    \hspace{2.5mm}
    \includegraphics[width=0.48\textwidth]{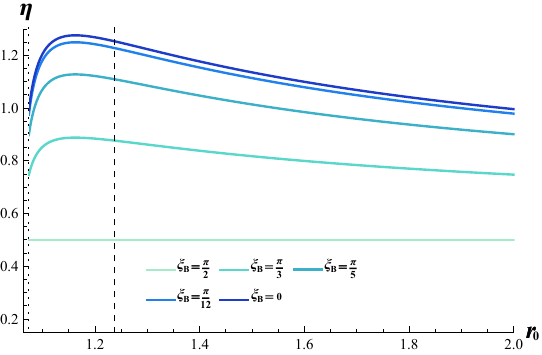}
    \caption{Energy-extraction efficiency $\eta$ as a function of the radius of reconnection point $r_0$ for multiple values of orientation angle $\xi_B$ with $\sigma_0=100$, $a=0.9$ (top panels) and $a=0.998$ (bottom panels) in the cases of the combined streamline (left panels) and the circular streamline (right panels). The dotted line represents the radius of the photon sphere while the dashed line represents the radius of the ISCO ($r_{\rm ms}>r_{\rm ergo}$ for $a=0.9$).}
    \label{fig:eta-xi}
\end{figure}

We plot the  energy-extraction efficiency as a function of $r_0$ for multiple values of orientation angle in Fig.~\ref{fig:eta-xi}, with $\sigma_0=100$, $a=0.9$ (top panels) and $a=0.998$ (bottom panels) in the cases of the combined (left panel) and the circular streamline (right panel). From the right panels, we see that $\eta$ solely decreases with the increase of the orientation angle in the case of circular streamline, which means the energy extraction becomes harder for a larger orientation angle, in consistency with previous results \cite{CA2021}. However, the situations for the plunging bulk plasma are different. One can see from the left panels that, in the plunging region, $\eta$ could become even larger with an increasing orientation angle. It indicates  that the energy extraction would be easier when the magnetic reconnection happens with a larger orientation angle if the bulk plasma plunges to the central black hole from the ISCO.

\begin{figure}
    \centering
    \hspace{-3mm}
    \includegraphics[width=0.48\textwidth]{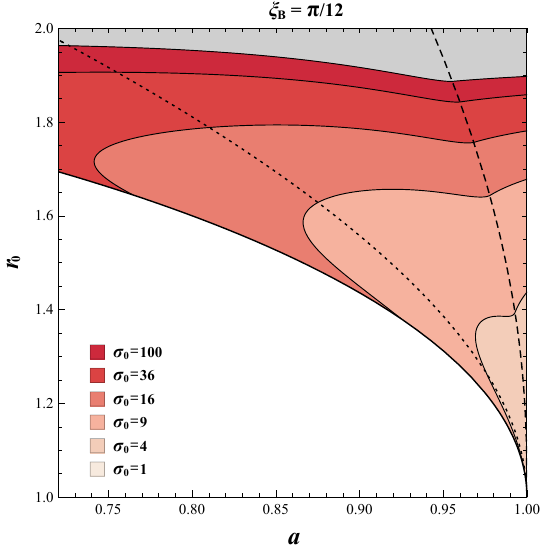}
    \hspace{3mm}
    \includegraphics[width=0.48\textwidth]{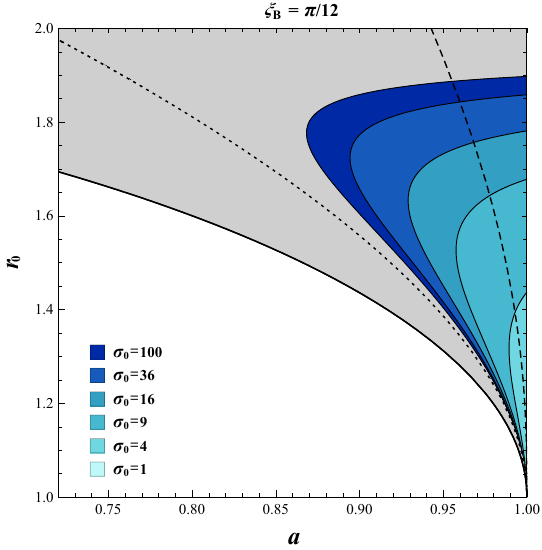} 
    \hspace{-3mm}
    \includegraphics[width=0.48\textwidth]{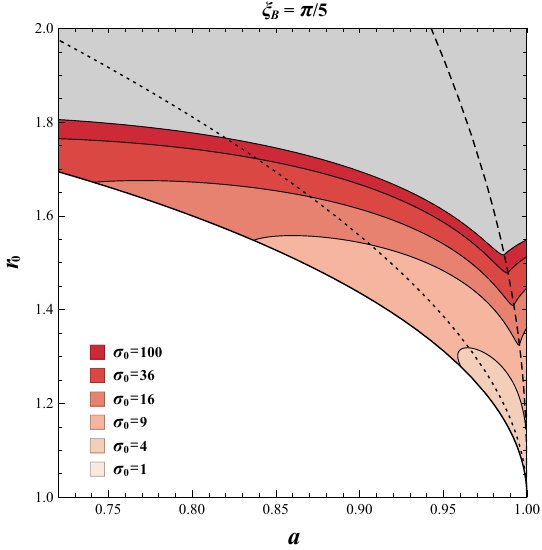}
    \hspace{3mm}
    \includegraphics[width=0.48\textwidth]{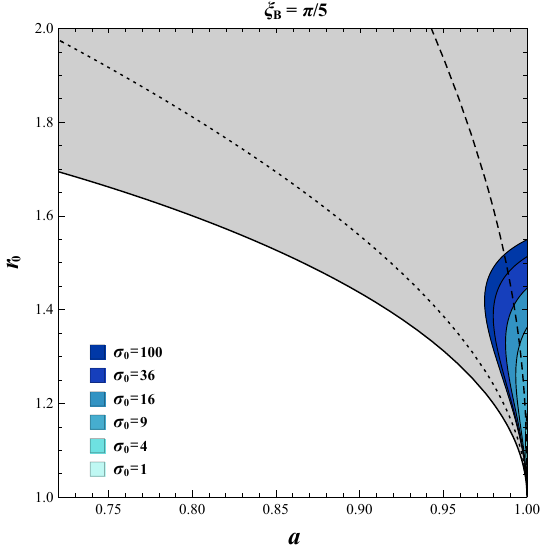}
    \caption{The $a-r_0$ planes in the case of the combined streamline (left panels) and the circular streamline (right panels), where the allowed regions for energy extraction ($\eta>0$), with $\xi_{\rm B}=\pi/12$ (top panels) and $\xi_{\rm B}=\pi/5$ (bottom panels), are colored in red (combined streamline) or blue (circular streamline) for multiple values of local magnetization $\sigma_0$. The solid, dotted and dashed black lines represent the radii of the event horizon, the photon sphere and the ISCO respectively.}
    \label{fig:r-a-12-5}
\end{figure}

We plot the allowed regions for energy extraction on $a-r_0$ planes for multiple values of local magnetization in Fig.~\ref{fig:r-a-12-5}, with $\xi_B=\pi/12$ and $\xi_B=\pi/5$. The allowed regions are vastly different between the cases of different streamlines. As already mentioned above, the  energy-extraction efficiency entirely becomes lower for larger orientation angle when the bulk plasma flows purely circularly. Hence the larger orientation angle makes the allowed region for energy extraction diminished (compare the right panels of Fig.~\ref{fig:r-a-12-5} to the right panel of Fig.~\ref{fig:r-a-0}). However, the picture would be different for the bulk plasma plunging from the ISCO: the larger orientation angle  makes the allowed region for energy extraction enlarged in the plunging region, as could be seen from the left panels in Fig.~\ref{fig:r-a-12-5}. For example,  the allowed regions for energy extraction in the case of combined streamline with both $\xi_B=\pi/12$ and $\xi_B=\pi/5$ can reach the region with  $a\lesssim 0.7$ when $\sigma\simeq 100$, which cannot be reached by the allowed region with $\xi_B=0$ (see left panel of Fig.~\ref{fig:r-a-0}) and for which the circular orbits disappear within the ergo-sphere.

\begin{figure}
    \centering
    \hspace{-3mm}
    \includegraphics[width=0.48\textwidth]{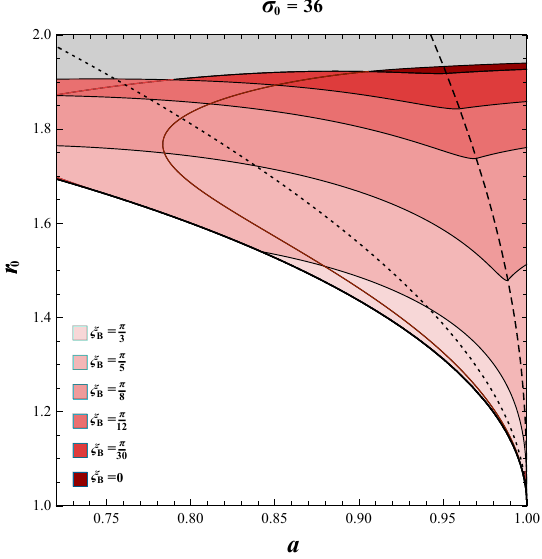}
    \hspace{3mm}
    \includegraphics[width=0.48\textwidth]{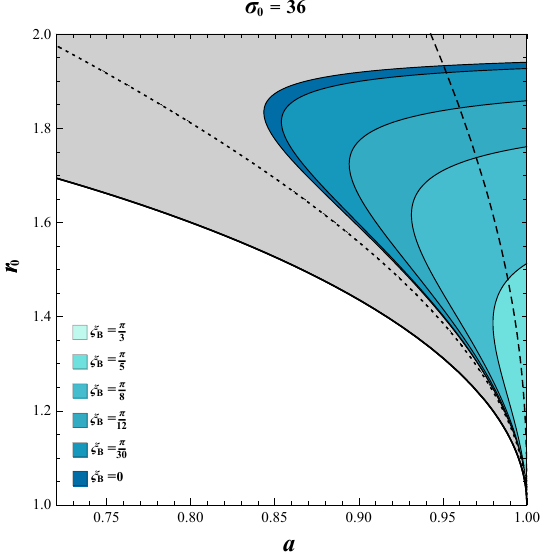}
    \caption{The $a-r_0$ planes in the cases of the combined streamline (left panels) and the circular streamline (right panels).The allowed regions for energy extraction ($\eta>0$), with $\sigma_0=36$, are colored in red (combined streamline) or blue (circular streamline) for multiple values of orientation angle $\xi_B$. The solid, dotted and dashed black lines represent the radii of the event horizon, the photon sphere and the ISCO, respectively. Some parts of the allowed region with $\xi_B=0$ and $\xi_B=\pi/30$ on the left panel are covered so we just plot their boundaries by lines with the same colors of the regions.}
    \label{fig:r-a-xi}
\end{figure}

One can see Fig.~\ref{fig:r-a-xi} to figure out how the allowed regions for energy extraction in $a-r_0$ planes vary with the change of orientation angle, where we choose $\sigma_0=36$ as an example. From the right panel, we find out that the allowed region becomes more and more diminished with the increase of orientation angle when the bulk plasma flows purely circularly, which is consistent with the result shown in Fig.~4 of Ref.~\cite{CA2021}. However, the left panel shows that, when the bulk plasma plunges within ISCO, the allowed region for energy extraction in the plunging region gradually changes the shape with the increase of orientation angle rather than purely shrinks. What is more, we know that photon sphere is a limitation of circular orbit, on which both $E$ and $L$ diverge and that is why the allowed regions for energy extraction in the case of circular streamline should be out of while smoothly approach to the photon sphere. However, there is no such a limitation for the plunging geodesic, on which $E$ and $L$ always keep constant. And that makes the allowed regions for energy extraction open regions in the case of combined streamline, except for the cutoffs on the event horizon.

\begin{figure}
    \centering
    \hspace{-3mm}
    \includegraphics[width=0.48\textwidth]{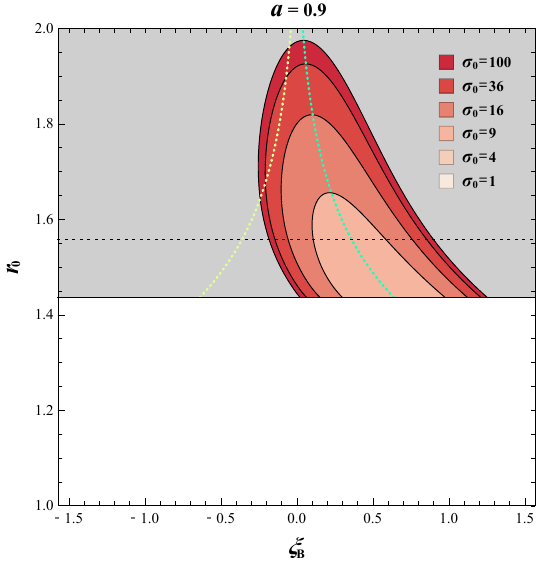}
    \hspace{3mm}
    \includegraphics[width=0.48\textwidth]{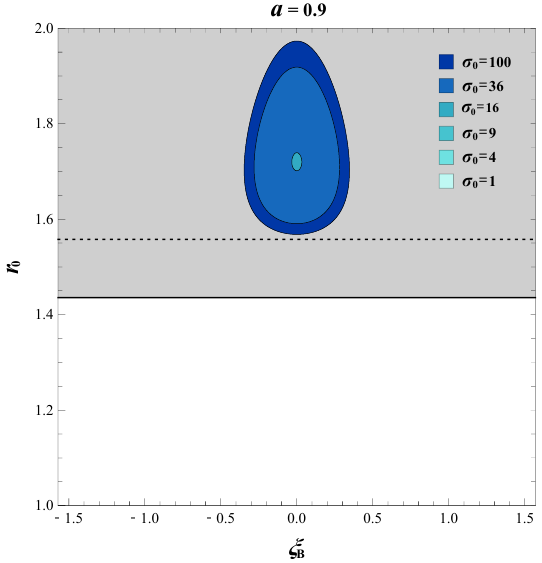} 
    \hspace{-3mm}
    \includegraphics[width=0.48\textwidth]{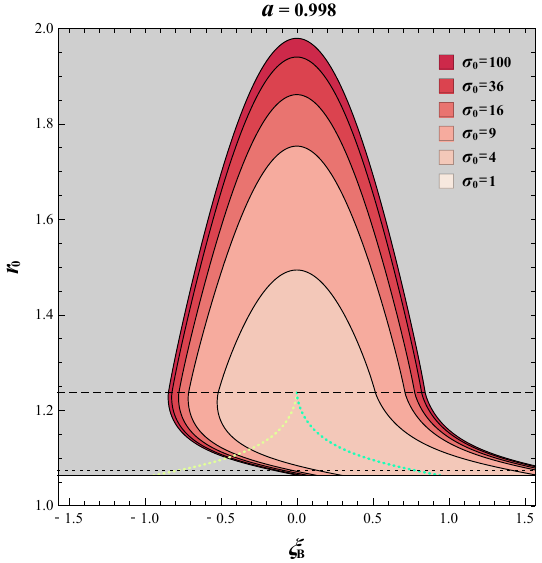}
    \hspace{3mm}
    \includegraphics[width=0.48\textwidth]{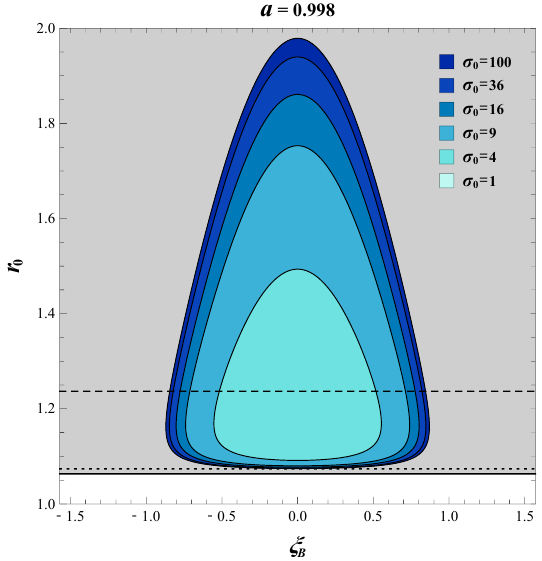}
    \caption{The $\xi_B-r_0$ planes for the combined streamline (left panels) and the circular streamline (right panels), where the allowed regions for energy extraction ($\eta>0$), in the case of $a=0.9$ (top panels) and $a=0.998$ (bottom panels), are colored in red (the combined streamline) or blue (the circular streamline) for multiple values of local magnetization $\sigma_0$. The solid, dotted and dashed black lines represent the radii of the event horizon, the photon sphere and the ISCO respectively ($r_{\rm ms}>r_{\rm ergo}$ for $a=0.9$). The dotted yellow lines on the left panels represent the orientation angles determined by ideal MHD condition. The cyan dotted lines on the left panels represent $\xi_{B,{\rm m}}$.}
    \label{fig:r-xi}
\end{figure}

We plot the allowed region for energy extraction on $\xi_B-r_0$ planes in the cases of $a=0.9$ (top panels) and $a=0.998$ (bottom panels) in Fig.~\ref{fig:r-xi}. The allowed regions on $\xi_B-r_0$ planes are visibly larger in the case of the combined streamline, especially when the black hole spin is not extreme (compare the top two panels) such that the plunging region is broad. Different from the allowed regions for energy extraction in the case of the circular streamline which is axisymmetric about $\xi_B=0$ solely, the allowed regions in the case of the combined streamline incline to the side of positive orientation angle in the plunging region, which indicates that suitable increase of orientation angle is favored to energy extraction if the bulk plasma plunges within the ISCO. It is consistent with the result when we compare the left panels of Fig.~\ref{fig:r-a-12-5} with the left panel of Fig.~\ref{fig:r-a-0}. 

Let us try to understand the phenomena shown in the above plots  through some basic analyses here. Revisiting Eq.~\eqref{eq:epsilon}, we can find out that, in order to produce a plasmoid with negative $\epsilon_-$ more easily, one would expect the value of
\begin{equation}
    \left(\hat{v}_s+\beta^{\phi}\frac{\hat{v}_s^{(\phi)}}{\hat{v}_s}\right)\cos\xi_{B}-\frac{\hat{v}_s^{(r)}}{\hat{\gamma}_s\hat{v}_s}\sin\xi_B
    \label{eq:EE_condition1}
\end{equation}
to be as large as possible, whose extreme value, when fixing the reconnection point, locates at
\begin{equation}
    \xi_B=\xi_{B,{\rm m}}\equiv \arctan\left(-\frac{\beta^{\phi}\hat{v}_s^{(r)}}{\hat{\gamma}_s\hat{v}_s^2+\beta^{\phi}\hat{\gamma}_s\hat{v}_s^{(\phi)}}\right).
    \label{eq:xi_m}
\end{equation}
In the case of circular streamline for which $\hat{v}_s^{(r)}=0$ such that $\xi_{B,{\rm m}}=0$, any non-zero orientation angle reduces the value of Eq.~\eqref{eq:EE_condition1}. However, if the bulk plasma plunges from the ISCO carrying negative $\hat{v}_s^{(r)}$, we could definitely find a positive value of Eq.~\eqref{eq:xi_m}. That is why suitably increasing the orientation angle could increase the  energy-extraction efficiency and broaden the allowed regions in $a-r_0$ planes in the case of the combined streamline as show on the left panels of Fig.~\ref{fig:eta-xi} and Fig.~\ref{fig:r-a-12-5}. That is also why in the case of the combined streamline the allowed regions in $\xi_B-r_0$ planes incline to the side of positive orientation angle as shown on the left panels of Fig.~\ref{fig:r-xi}. We also plot $\xi_{B,{\rm m}}$ by dotted cyan lines in the plunging regions on the left panels of Fig.~\ref{fig:r-xi}. The tendencies of the allowed regions for energy extraction inclining to the side of positive orientation angle, when the bulk plasma plunges, almost follow the trends of $\xi_B=\xi_{B,{\rm m}}$ (not exact because of the last term in Eq.~\eqref{eq:epsilon}), which becomes more obvious in the high magnetization limit (in which the last term in Eq.~\eqref{eq:epsilon} vanishes). As a consequence, we may call $\xi_{B,{\rm m}}$ the best orientation angle in the high magnetization limit. 

Intuitively, one may expect the plasmoid to be ejected azimuthally symmetric (perpendicular to $\hat{e}_{(r)}$) to maximize the energy-extraction efficiency. In this sense, $\xi_B=0$ is the best choice in the case of the circular streamline. When the bulk plasma plunges from the ISCO, the plasmoids would be ejected azimuthally symmetric when $u_{\pm}^{\mu}-\gamma_{\rm out}e_{[0],0}^{\mu}$ is parallel (or anti-parallel) to $\hat{v}_s^{(\phi)}e_{[3]}^{\mu}-\hat{\gamma}_s\hat{v}_s^{(r)}e_{[1]}^{\mu}$, for which the orientation angle should be:
\begin{equation}
    \xi_B=\arctan\left(-\frac{\hat{\gamma}_s\hat{v}_s^{(r)}}{\hat{v}_s^{(\phi)}}\right),
\end{equation}
which is not identical to $\xi_{B,{\rm m}}$. In this case, the best orientation angle for energy extraction when the bulk plasma plunges would not makes the plasmoids ejected in an azimuthally symmetric way.

\begin{figure}
    \centering
    \includegraphics[width=0.7\textwidth]{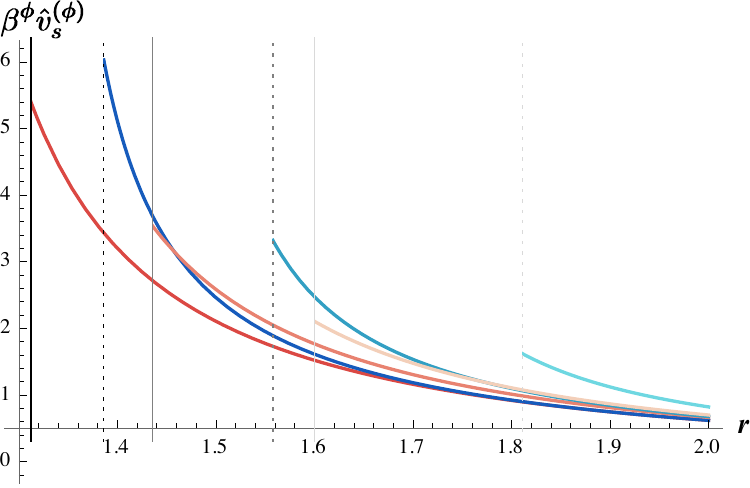}
    \caption{$\beta^{\phi}\hat{v}_s^{(\phi)}$ as a function of radius with $a=0.95$, 0.9 and 0.8 (deeper to lighter color) in the cases of combined (red lines) and circular (blue lines) streamlines. The radii of event horizon and photon sphere are represented by solid and dotted line respectively.}
    \label{fig:beta_vs}
\end{figure}

On the other hand,  one can see from \eqref{eq:xi_m} that a smaller value of
\begin{equation}
    \beta^{\phi}\hat{v}_s^{(\phi)}=\frac{\omega^{\phi}L}{E-\omega^{\phi}L}
    \label{eq:EE_condition2}
\end{equation}
is favored for energy extraction. The radius dependence of $\beta^{\phi}\hat{v}_s^{(\phi)}$ for multiple values of black hole spin in the cases of both the combined and the circular streamlines are plotted in Fig.~\ref{fig:beta_vs}, from which one can visibly figure out that the value of $\beta^{\phi}\hat{v}_s^{(\phi)}$ are generally larger in the circular streamline than in the combined streamline. In this sense, as a conclusion, the energy extraction via magnetic reconnection in bulk plasma with circular streamline is generally harder, which explains the lower values of the  energy-extraction efficiency shown in Fig.~\ref{fig:eta-ck} and more restrictive the allowed regions for energy extraction shown in Fig.~\ref{fig:r-a-0}. Notice that $\beta^{\phi}\hat{v}_s^{(\phi)}$ is independent of the magnetic reconnection process. It purely determined by the motion of the bulk plasma and the spacetime geometry.

The dotted yellow lines on the left panels of Fig.~\ref{fig:r-xi} represent the orientation angles determined by ideal MHD condition in the plunging region \cite{Work0,Hou:2023bep,Zhang:2024lsf}. As we can see, only small parts of the yellow lines lay on the allowed region for energy extraction, which means that, although the allowed regions are large in the case of the combined streamline, the energy extraction in the plunging region would still not be highly probable if the orientation angle is determined by the large-scale magnetic configuration obeying ideal MHD condition, which always results in negative orientation angles. This is consistent with the results in Ref.~\cite{Work0}.

\subsection{Covering factor for energy extraction}
\label{sec:chi}

In a background of fixed black hole spin, magnetization and streamline, we may define the following quantity 
\begin{equation}
    \chi\equiv \frac{S_{\rm EE}}{\pi\left(r_{\rm ergo}-r_{\rm EH}\right)}
    \label{eq:chi}
\end{equation}
to be the covering factor for energy extraction from a rotating black hole via magnetic reconnection, where $S_{\rm EE}$ is the ``area" of the allowed region for energy extraction in $\xi_B-r_0$ plane. This quantity reflects how probable the energy extraction could be realized via magnetic reconnection occurring within the ergo region of a rotating black hole, by assuming the occurrence of magnetic reconnection in any direction on any position of the equatorial plane within the ergo-sphere is equally probable. Although this assumption is generally unrealistic since the local direction along which the magnetic reconnection occurs is highly dependent on the large scale configuration of magnetic field. Moreover, the magnetization of accretion flow in the ergo region could not be uniformly distributed generally, which subsequently makes the probabilities unequal that the magnetic reconnection occurs in different positions. However, it is not bad for us to analyze the covering factor as a primitive trial to judge the capability of an accretion system on extracting energy from the central black hole via magnetic reconnection.

\begin{figure}
    \centering
    \includegraphics[width=0.7\textwidth]{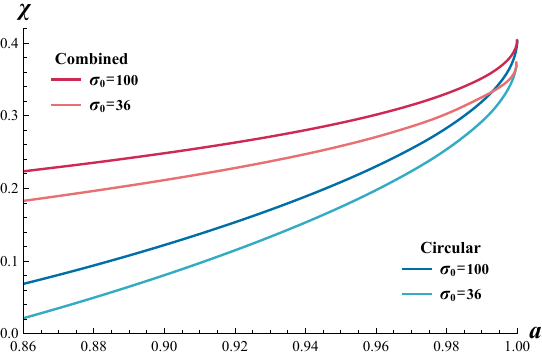}
    \caption{The covering factor $\chi$ defined in Eq.~\eqref{eq:chi} as a function of black hole spin in the cases of the combined (red lines) and the circular (blue lines) streamlines with $\sigma_0=36$ (light color) and $\sigma_0=100$ (deep color).}
    \label{fig:chi}
\end{figure}

In Fig.~\ref{fig:chi}, we plot the covering factors as functions of black hole spin for the two kinds of streamlines with $\sigma_0=36$ and $\sigma_0=100$. We can see that $\chi$ in the case of the combined streamline are generally much larger than those of the circular streamline. In this sense, one could opine that the plunging bulk plasma would be more capable of extracting energy from a rotating black hole via magnetic reconnection than the circularly flowing bulk plasma, if they are equally magnetized. As the black hole turns to be near extreme, the difference between the covering factors become less and near to zero, due to the fact that $r_{\rm ms}$ converges to $r_{\rm EH}$ and the plunging region shrinks gradually and disappears finally.

\begin{figure}
    \centering
    \hspace{-2.5mm}
    \includegraphics[width=0.48\textwidth]{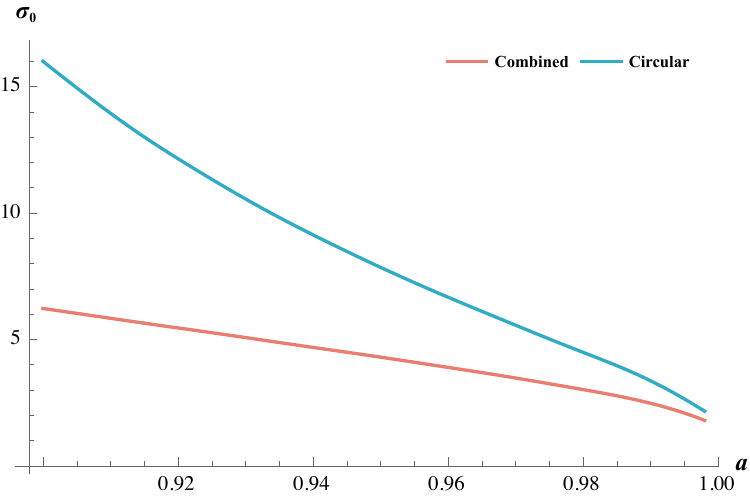}
    \hspace{2.5mm}
    \includegraphics[width=0.48\textwidth]{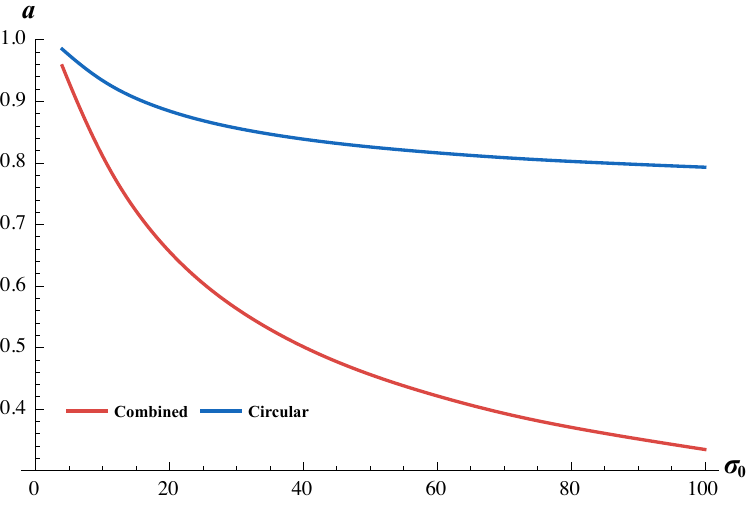}
    \caption{Minimal local-magnetization $\sigma_{\rm 0,min}$ for energy extraction as a function of black hole spin ranging from 0.9 to 0.998 (left panel) and minimal black hole spin $a_{\rm min}$ for energy extraction as a function of local magnetization ranging from 5 to 100 (right panel) for the combined (red lines) and the circular (blue lines) streamlines.}
    \label{fig:min}
\end{figure}

One could determine the minimal local magnetization $\sigma_{0,{\rm min}}$ for energy extraction by solving $\chi=0$ for a fixed spin, as shown on the left panel of Fig.~\ref{fig:min} where $a$ ranges from 0.9 to 0.998. On the contrary, a minimal spin $a_{\rm min}$ of black hole for energy extraction could be determined in this way for a fixed local magnetization as well, as shown on the right panel of Fig.~\ref{fig:min} where $\sigma_0$ ranges from 5 to 100. One can see from Fig.~\ref{fig:min} that the energy extraction via magnetic reconnection occurring in a circularly flowing bulk plasma is much more restrictive. Although the circular orbits could reach the ergo-sphere as long as $a\gtrsim 0.7$, it requires $a\gtrsim 0.8$ for energy extraction in the case of circular streamline even in a high magnetization ($\sigma\simeq 100$). On the other hand, the energy extraction via magnetic reconnection in the plunging plasma is less restrictive: It can happen even for a slowly rotating black hole if the magnetization is strong enough, as shown on the right panel of Fig.~\ref{fig:min}.


\section{Summary and discussions}
\label{sec:sum}

In this work, we revisited  the energy extraction via magnetic reconnection, specifically via the Comisso-Asenjo process, from a rotating black hole. Different from most of the relevant studies in the literatures \cite{CA2021}, which focused on the circularly flowing bulk plasmas even within the ISCO, we paid more attention to the combined streamline scenario, in which the bulk plasma flows circularly at the initial stage but then plunges from the ISCO to the event horizon. Moreover, we relaxed the restriction on the orientation angle that was fixed in Ref.~\cite{Work0} by imposing ideal MHD condition, and we allowed it to be a free parameter, considering the fact that the realistic magnetic configuration could be unpredictable.   


We compared the efficiencies of energy extraction and the allowed regions for energy extraction in $a-r_0$ plane with zero orientation angle between the bulk plasmas with two kinds of streamlines. We figured out that the bulk plasma with the combined streamline generally have higher  energy-extraction efficiency, especially for the non-extreme black hole, and have larger allowed region as well.

Furthermore, we analyzed the influence of orientation angle on the energy extraction. We notice that in the circularly flowing bulk plasma, larger orientation angles generally make the energy extraction harder, which is reflected on the decreasing  efficiencies and the shrinking  allowed region for energy extraction in $a-r_0$ planes. However, the situations in the bulk plasma with a combined streamline are quite different. An appropriate increase of the orientation angle in this case would make the energy extraction even easier in the plunging region, which is reflected by the increases of efficiencies and the expanded allowed region in $a-r_0$ planes. We figured out the best orientation angle $\xi_{B,{\rm m}}$ in the high magnetization limit by minimizing $\epsilon_-$ expressed in Eq.~\eqref{eq:epsilon}, showing that for the plunging bulk plasma, the plasmoids are not ejected azimuthally symmetric with the best orientation angle, in contrary to the intuitive expectation.

We plotted the allowed regions for the energy extraction in $\xi_B-r_0$ planes, in order to  see clearly the dependence of the energy extraction on the orientation angle for the bulk plasma with two kinds of streamlines. We found out that, different from the the allowed regions for energy extraction in the case of circularly flowing bulk plasma which is symmetric about $\xi_B=0$ on $\xi_B-r_0$ planes, the allowed regions for the bulk plasma with the combined streamline generally are inclined to the side of positive orientation angles. On $\xi_B-r_0$ planes, we also plotted the $\xi_{B,{\rm m}}$ and the orientation angle determined by the ideal MHD condition in the plunging region. It could be seen from the plots that the allowed regions for energy extraction tend  towards the sides of positive orientation angle, in the case of the plunging bulk plasma, almost follow the trends of $\xi_B=\xi_{B,{\rm m}}$, especially in the high magnetization limit. 

We tried to primitively investigate the energy extraction probability by defining the covering factor $\chi$, which is the ratio of the area of the allowed region to the total area of $\xi_B-r_0$ plane. The quantity $\chi$ depends on central black hole spin and the properties of bulk plasma, and it reflects the capability of the bulk plasma in extracting energy from the central black hole, under the assumption that magnetic reconnection would be equally probable to occur along any direction on any position within the ergo-sphere. By comparing the covering factors, we argued that the plunging bulk plasma is generally more capable of extracting energy than the circularly flowing one. By solving $\chi=0$, we determined the minimal local magnetization for energy extraction as a function of black hole spin. In a similar way, we read the minimal spin of the black hole as a function of local magnetization, from which we found out that the energy extraction is feasible even from a slowly rotating black hole if the bulk plasma plunges within the ISCO.


It should be noticed here that in this work we did not take into account of  the escaping condition for an efficient energy extraction \cite{Work0}, which requires the ejected plasmaid with $\epsilon_+$ not to be attracted down to the event horizon. We believe that this defect would not be essential to our analysis. On one hand, the results in Ref.~\cite{Work0} showed that the escaping conditions could exclude only a little part of the allowed region for energy extraction in $a-r_0$ plane for the plunging bulk plasma even though the orientation angles determined by ideal MHD condition are always negative. On the other hand, based on our results, the allowed region for the energy extraction in the case of the plunging bulk plasma tends to the side of positive orientation angles, with which it is easier for the ejected plasmoid to escape. Consequently, the influence of escaping condition should have insignificant influence to our results. 

Moreover, recent studies show that the spacetime curvature would have great impacts on the properties of magnetic reconnection \cite{CA2017,Comisso:2018ark,Fan:2024fcy,Shen:2024plw}, which have to be taken into account when considering the magnetic reconnection occurring near a rotating black hole. However, Ref.~\cite{CA2017} and Ref.~\cite{Comisso:2018ark} only considered two simple cases that the current sheet in Sweet-Parker configuration were posited azimuthally and radially. Though the authors in Ref.~\cite{Fan:2024fcy} extended the analyses to a more general case, they did not analyze how the outflow speed is affected. The gravitational effects on properties of magnetic reconnection process need further exploration.

The magnetic reconnection is believed to occur frequently in the accretion flow around massive black holes, which has been partly supported already by numerical simulations \cite{Yuan2024-1,Ripperda:2021zpn}. More importantly, the hot flows generated by magnetic reconnection in accretion disk may trigger the emissions. These emissions carry remarkable information on the spacetime and the magnetic configurations within a few gravitational radii from the central black hole in the spectra and the polarizations \cite{Jia:2023iup,EHT:2024nwx,GRAVITY:2018det,GRAVITY:2020lpa}. With the development of observational precision, it is worth looking forward to detect the signals from the plasmoids produced by magnetic reconnection in the near future \cite{Emami:2022ydq}, from which the probable energy extraction may be identified from the observation.

\section*{Acknowledgement}

We highly appreciate the help from Yehui Hou and Junyi Li. We thank the anonymous referee for valuable comments and constructive suggestions. The work is partly supported by NSFC Grant Nos. 12275004.

\appendix

\bibliographystyle{utphys}
\bibliography{references}

\end{document}